\documentstyle[12pt,epsf]{article}
\input psfig
\begin{document}
\title{$\Lambda$-inflation and CMB anisotropy}
\author{V.N.Lukash and E.V.Mikheeva\\
Profsoyuznaya 84/32, Moscow 117810, Russia}
\maketitle


\centerline{\bf Abstract}
We explore a broad class of three-parameter inflationary models, 
called the $\Lambda$-inflation, and its observational 
predictions: high abundance of cosmic gravitational waves 
consistent with the Harrison-Zel'dovich spectrum of primordial 
cosmological perturbations, the non-power-law wing-like spectrum 
of matter density perturbations, high efficiency of these models 
to meet current observational tests, and others. We show that a 
parity contribution of the gravitational waves and adiabatic 
density perturbations into the large-scale temperature 
anisotropy, T/S $\sim 1$, is a common feature of 
$\Lambda$-inflation; the maximum values of T/S (basically not 
larger than 10) are reached in models where (i) the local 
spectrum shape of density perturbations is flat or slightly red 
($n_S{}_\sim^< 1$), and (ii) the residual potential energy of the
inflaton is near the GUT scale ($V_0^{\frac{1}{4}} \sim 10^{16}
GeV$). The conditions to find large T/S in the paradigm of cosmic 
inflation and the relationship of T/S to the ratio of the power 
spectra, $r$, and to the inflationary $\gamma$ and Hubble 
parameters, are discussed. We argue that a simple estimate, 
T/S$\simeq 3r\simeq 12\gamma \simeq \left(\frac{H}{6\times 
10^{13}{\rm GeV}}\right)^2$, is true for most known inflationary 
solutions and allows to relate straightforwardly the important 
parameters of observational and physical cosmology.  

\newpage

\section{Introduction}

The situation in physical cosmology is currently governed by experiment
(observations) which made an increasing progress for the recent years.
However, the theory of formation of {\it Large Scale Structure} in the
Universe leaves something to be desired while a progress is still there: 
the simplest versions of the dynamical {\it Dark Matter} are discarded 
(e.g. sHDM, sCDM, cosmic strings), the cosmological model has become
multiparametrer ($\Omega_{{\rm M}}$, $\Omega_{\Lambda}$, $\Omega_b$, $h$, 
$n_{{\rm S}}$, T/S, etc.) which hints on a complex nature of the dark matter
in the Universe. Hopefully, the ongoing and oncoming measurements of the CMB
anisotropy (both ground and space based) as well as the development of
median and low $z$ observations will fix the DM/LSS model of the Universe by
a few per cent in the nearest future.

A theory of the very early Universe which meets most predictions 
and observational tests is inflation. It prophesys small Gaussian 
{\it Cosmological Density Perturbations} (the {\it Scalar} adiabatic mode)
responsible for the LSS formation in the observable Universe. The ultimate
goal here would be the reconstruction of DM parameters and CDP power
spectrum directly from observational data (LSS {\it vs} $\Delta T/T$).

A drama put in the basis of cosmic inflation is that it provides also a
general ground for the fundamental production of {\it Cosmic Gravitational
Waves} (the {\it Tensor} mode) which should contribute along with the S-mode
into the $\Delta T/T$ anisotropy at large angular scale\footnote{
Obviously, all three modes of the perturbations of gravitational field --
scalar, vector and tensor (see [1]) -- induce the CMB anisortopy through the
SW-effect [2]. However, most of the inflationary models considered by now
are based on scalar inflaton fields which cannot be a source for the vector
mode. A general physical reason for the production of the T and S
perturbations in the expanding Universe is the {\it parametric amplification
effect} [3], [4]: the spontaneous creation of quantum physical fields in a
non-stationary gravitational background.}. Hence, a principal question on
the way to the S-spectrum restoration remains the T/S problem -- the
fraction of the variance of the CMB temperature fluctuations on 10 degrees
of arc generated by the CGWs: 
\begin{equation}
\left(\Delta T/T\right)^2 \vert_{10^0} = {\rm S} + {\rm T}.
\end{equation}
Observational separation between the modes is postponed by the time when
polarization measurements of the CMB anisotropy will be available (which
require the detector sensitivity ${}_{\sim}^{<} 1\mu$K). Today, we can
investigate the T/S problem theoretically.

A common suggestion created by the {\it Chaotic Inflation} [5], that 'T/S 
{\it is usually small} (T/S ${}_{\sim}^{<} 0.1$)', stems actually from a
very specific property of the CI model (it inflates only at high values of
the inflaton, $\varphi >1 $). However, in general this is not true: any
inflation produces inevitably {\it both} pertubation modes, the ratio
between them is not limited by unity and sticks to the parameters of a 
given model\footnote{There is no fundamental theorem restricting T/S 
relative to the unity: the inflationary requirement, $\gamma\equiv - 
\dot{H}/H^2 < 1$, imposes only a wide constraint, T/S ${}_{\sim}^< 10$, 
obviously insufficient to discriminate the T-mode in the cosmological 
context (see eqs.(2),(5)).}. Nevertheless, people often relate this 
T/S-CI feature to another basic property of the chaotic inflation with 
a smooth inflaton potential $V(\varphi)$ -- the {\it Harrison-Zel'dovich} 
S-spectrum ($n_S\simeq 1$).

Such a mythological statement that 'T/S {\it is small when} $n_S\simeq 1$',
has even been strenthened by the power-law inflation [6], [7] which has
displayed that T/S may become large only at the expense of the rejection
from the HZ-spectrum in S-mode: T/S ${}^{>}_\sim 1 $ when $n_S\leq 0.8$;
obviously, {\it vice versa}, when $n_S\rightarrow 1$, the T/S tends to zero
in a total accordance with the previous CI-assertion. The analytic
approximation for T/S found in this model looks eventually universal for any
inflationary dynamics when related to the T-spectrum slope index (estimated
in the appropriate scale $k_{COBE}\sim 10^{-3}h/$Mpc)\footnote{It is just 
because the CGW spectrum created in {\it any} inflation is
intrinsically akin to the evolution of the Hubble factor at the horizon
crossing time: $n_T \simeq - 2\gamma < 0$. Notice the T-spectrum stays
always red in the minimally coupled gravity since a systematic decrease in
time of the Hubble factor, cf.eq.(12).}, 
\begin{equation}
\frac {{\rm T}}{{\rm S}} \simeq -6n_T \simeq 12\gamma.
\end{equation}

Since a case for the {\it red} S-spectra suggested by power-law inflation 
($n_S < 1$) has confirmed the above statement of the T/S smallness for
HZ-CDPs, we are facing to check the opposite situation -- a case for models
where the {\it blue} spectra ($n_S > 1$) are allowed and the T/S there. An
example of the blue S-spectrum is provided by (i) the two-field hybrid
inflation [8], [9], [10], [11] for a certain range of the model parameters,
(ii) a single massive inflaton [12] ($V=V_0 + m^2\varphi^2/2$), and (iii)
that producing power-law S-spectrum [13], [14], [15]. However, the problem
of blue S-spectra is more generic and requires its full investigation. In
this paper we present such analysis for the case of a {\it single} inflaton
field.

Below, we start considering the inflationary requirements for the production
of blue S-spectra. We introduce a simple natural model of such an inflation
with one scalar $\varphi$ field which we call the {\it $\Lambda$-inflation}.
It proceeds at any values of the inflaton and generates a typical feature in
the S-spectrum: a blue branch in short wavelengths (small $\varphi$ values)
and a red one in large wavelengths (high $\varphi$ values). Between these
two asymptotics the broad-scale transient spectrum region is settled down
where the 'T/S {\it is close to its highest value (generally not more than
10) when the} S {\it spectrum (or the joint} S+T {\it metric fluctuation
spectrum) is essentially HZ one}'. Futher on, we analyse physical reasons
for the latter generic statement (CI is the measure zero in the family of 
$\Lambda$-inflation models) and its place in the inflationary paradigm.
Surprisingly, the phenomena of large T/S and blue S-spectra are two totally
disconnected problems: both are realized in $\Lambda$-inflation but at
different scales and field values. The large T/S is produced where inflation
proceeds only marginally (the subunity $\gamma$-values) which occurs near 
$\varphi\sim 1$ where the S-spectrum tilt is slightly red, $n_S{}_\sim^< 1$.
On the contrary, the blueness ($n_S > 1$) is gained for $\varphi \ll 1$ and
has thus a different physical origin. We conclude by discussing the
necessary and sufficient conditions for obtaining large T/S from inflation,
and argue for a general estimate of T/S based on eq.(2).

\section{The $\Lambda$-Inflation}

We are looking for the simplest way to get a blue-kind spectrum of density
perturbations generated at inflation driven by one scalar field $\varphi$.

The minimal coupling of $\varphi$ to geometry is given by the action ($c =
\hbar = 8\pi G = 1$): 
\begin{equation}
W \left[\varphi, g^{ik}\right] = \int \left(L - \frac 12 R\right) \sqrt{-g}
\; d^4 x
\end{equation}
where $g_{ik}$ and $R_{ik}$ are the metric and Ricci tensors respectively,
with the signature $(+ - - -)$, $g = det (g_{ik})$, and $R \equiv R^{i}_{i}$.
The field Lagrangian is an arbitrary function of two scalars, 
\begin{equation}
L = L\left(w, \varphi\right),
\end{equation}
where $w^{2} = \varphi_{,i} \varphi^{,i}$ is the kinetic term of 
$\varphi$-field.

Actually, the latter can be simplified at inflation. Indeed, the
inflationary condition (taken in the locally flat Friedmann geometry), 
\begin{equation}
\gamma \equiv - \frac{\dot{H}}{H^{2}} = \frac{3(\rho + p)}{2\rho} = \frac{3
w^{2} M^{2}}{2(w^{2} M^{2} - L)} < 1,
\end{equation}
implies generally that 
\[
w^2 M^{2} \equiv \frac{\partial L}{\partial (\ln w)} < - 2L, 
\]
just telling us on the validity of the Taylor-decomposition of (4) over
small $w^{2}$: 
\[
L=L(0,\varphi)+\frac{1}{2}w^{2}M^{2}(0,\varphi)+0(w^{4}), 
\]
where $\rho\equiv w^2M^2-L$ and $p\equiv L$ are comoving density and
pressure of $\varphi$-field, $H=\frac{g_{,i}\varphi^{,i}}{6w g}$ is the
local Hubble factor. After redefining the field by a new one, 
\[
\varphi \Rightarrow \int M (0, \varphi) d \varphi, 
\]
we come to a standard form for the Lagrangian density at inflation which is
assumed further on: 
\begin{equation}
L = - V (\varphi) + \frac{w^{2}}{2}.
\end{equation}
Here $V = V(\varphi)$ is the potential energy of $\varphi$-field.

A simple guess on the condition necessary to arrange inflation with a blue
S-spectrum arises when we address an example of the slow-roll approximation.
Under this approach the spectrum of created scalar perturbations $q_{k}$ is
straightforwardly related to the inflaton potential $V(\varphi) \simeq 3
H^{2}$ at the horizon-crossing: 
\begin{equation}
q_{k}\simeq \frac{H}{2\pi\sqrt{2\gamma}} = \frac{H^2}{4\pi
H^{\prime}_{\varphi}},\;\;\;\; k=aH=\dot{a},
\end{equation}
where $a$ is the scale factor and dot denotes the time derivative. The wave
number $k$ increases with time as $a$ grows faster than $H^{-1}$ in any
inflationary expansion (see eq.(5)): 
\[
\left(\ln\left(aH\right)\right)^{.}=\left(1-\gamma\right)H>0. 
\]
Eq.(7) prompts evidently: while decreasing $H^\prime_{\varphi}$ with 
$k>k_{cr}$, one gains the power on short scales and, thus, realizes the blue
spectrum slope.

Without loss of generality, we will assume that $V(\varphi)$ is a function
growing with $\varphi(>0)$ and getting its local minimum at $\varphi = 0$.
It means that during the process of inflation $\varphi$-field evolves to
smaller values. Hence, the necessary condition for a blue spectrum could be
any way of flattenning the potential shape at smaller $\varphi< \varphi_{cr}$
to provide for a rise of $H^{2}/H^\prime_{\varphi}$ and keeping the
inflation still on ($H^{\prime}_{\varphi} < H/ \sqrt{2}$, cf. eqs.(5), (6)): 
\begin{equation}
1-n_S\simeq \frac{\gamma}{H} \left(\frac{H^{2}}{H^{\prime}_{ \varphi}}
\right)^{\prime}_{\varphi} < 0.
\end{equation}

The latter equation leads to a broad-brush requirement of the positive
potential energy at the local minimum point of $\varphi$-field: 
\begin{equation}
V_{0}\equiv V(0) > 0,
\end{equation}
which displays the existence of the effective $\Lambda$-term during the
period of inflation dominated by the residual (constant) potential energy: 
\begin{equation}
V(\varphi<\varphi_{cr})\simeq V_{0}\equiv\Lambda\equiv 3H_0^2,
\end{equation}
where the characteristic value $\varphi_{cr}$ is determined as follows
\footnote{In most applications $\varphi_{cr}\sim 1$, see eq.(32).}: 
\begin{equation}
V (\varphi_{cr}) = 2 V_{0}.
\end{equation}

This appearance of the {\it de Sitter}-type inflation (for $\varphi <
\varphi_{cr}$) results in a drastic difference with CI which has eventually
assumed that $V_{0}=0$ just making the inflation at small $\varphi$ in
principal impossible. Obviously, the latter hypothesis on vanishing the
potential energy at $\varphi=0$ has reduced the CI-model to a very partial
case (from the point of view of eq.(9)) restricting the inflation dynamics
by only high values of the inflaton ($\varphi >1$). So, we may conclude that
the $\Lambda$-inflation based on eq.(9) presents a general class of the
fundamental inflationary models. In this sense they are more natural models
(the CI being of the measure zero in $V_{0}$-parameter) allowing the
inflation also at small $\varphi$-values (less than the Planckian one).

Summarising, we see that under condition (9) we have two qualitatively
different stages of the inflationary dynamics separated by 
$\varphi\sim\varphi_{cr}$. We will call them:

\begin{itemize}
\item  the CI stage ($\varphi {}_{\sim }^{>}\varphi_{cr}$), where the
evolution is not influenced by the $\Lambda $-term and looks essentially
like in standard chaotic inflation, and

\item  the dS stage ($\varphi <\varphi _{cr}$) predominated by the 
$V_0$-constant.
\end{itemize}

The completion of the full inflation in this model is related to $V_{0}$
-decay which is supposed to happen at some $\varphi^{\ast} < \varphi_{cr}$
\footnote{We do not discuss here possible mechanisms for such metastability 
(it may be the coupling to other physical fields, a way of double- or 
platoo-like inflations, etc.) and take the $\varphi^{\ast}$ value as an 
arbitrary parameter of our model (allowing to recalculate $k_{cr}$ in Mpc). 
Mind that in CI $\varphi^{\ast} \simeq \varphi_{cr}$.}. So, we deel with the
three-parameter model $(V_0, \varphi_{cr}, \varphi^{\ast})$ starting as CI 
($\varphi >\varphi_{cr}$) and processing by dS-inflation at small $\varphi$ 
($\varphi^{\ast} < \varphi < \varphi_{cr}$).

As we know from the CI theory smooth $V$-potentials create generally the 
{\it red} $q_{k}$-spectra ($n_S<1$ for $\varphi > \varphi_{cr}$). On the
other hand, eq.(9) provides physical grounds for the {\it blue} spectra
generated at dS period ($n_S>1$, cf. eq.(8)). Recall for comparison, that
the spectrum of gravitational waves produced at any inflationary regim is
given by the universal formula (here both polarizations are taken into
account): 
\begin{equation}
h_k=\frac H{\pi\sqrt 2}, \;\;\;\; k = a H,
\end{equation}
which generally ensures the red-like T-spectra as $H$ decreases in time for 
$\rho+p>0$: $n_T = -2\gamma <0$ (see eq.(5)).

A trivial way to maintain eq.(9) is the introduction of an additive 
$\Lambda$-term in the inflation potential. Keeping in mind only the simplest
dynamical terms we easily come to a trivial and rather general potential
form: 
\begin{equation}
V=V_0+\frac 12 m^2\varphi^2+\frac 14\lambda\varphi^4,
\end{equation}
which may also be understood as a decomposition of $V(\varphi)$ over small 
$\varphi$. Here, such decomposition is a reasonable approach since the
inflation proceeds to small $\varphi \rightarrow 0$. Obviously, eq.(11) can
be explicitely reversed in this case: 
\begin{equation}
\varphi^2_{cr}=\frac{4 V_0}{m^2+\sqrt{m^4+4\lambda V_0}}.
\end{equation}
Also, we will use later the power-law potential 
\begin{equation}
V=V_{0}+\frac{\lambda_{\kappa}}{\kappa}\varphi^{\kappa}=
V_{0}\left(1+y^{\kappa}\right),
\end{equation}
where $\kappa$ and $\lambda_{\kappa}$ are positive numbers ($\kappa\ge 2$, 
$\lambda_2\equiv m$, $\lambda_4\equiv\lambda$), $\varphi_{cr}^{\kappa}=\kappa
V_0/\lambda_{\kappa}$, and $y=\varphi/\varphi_{cr}$.

Let us turn to the evolution and spectral properties of $\Lambda$-inflation
models.

\section{The background model}

Below, we consider dynamics under the condition (9).

The background geometry is classical employing the 6-parametric Friedmann
group: 
\begin{equation}
ds^{2}=dt^{2}-a^{2}d\vec{x}^{2}=a^{2}(d\eta^{2}-d\vec{x}^{2}),
\end{equation}
The functions of time $a$ and $\varphi$ are found either from the Einstein
equations: 
\begin{equation}
H^{2} = \frac{1}{3} V + \frac{1}{6} \dot{\varphi}^{2},
\end{equation}
\begin{equation}
\dot{H} = - \frac{1}{2} \dot{\varphi}^{2},
\end{equation}
or equivalently, from the $\varphi$-field equation (with $H$ taken from
eq.(17)): 
\begin{equation}
\ddot{\varphi} + 3 H \dot{\varphi} + V^{\prime}_{\varphi} = 0.
\end{equation}
Coming to the dimensionless quantities, 
\[
h \equiv \frac{H}{H_{0}}, \;\;\;\; v = v(y) \equiv \left(\frac{V}{V_{0}}
\right)^{1/2},
\]
\begin{equation}
y \equiv \frac{\varphi}{\varphi_{cr}}, \;\;\;\; x \equiv H_0 \left(t -
t_{cr}\right), \;\;\;\; \epsilon \equiv \frac {2}{\varphi_{cr}},
\end{equation}
we can derive the first-order-equation for the function $h=h(y)$\footnote{
Hereafter, the prime/dot will denote the derivative over $y/x$, i.e. the
normalized $\varphi/t$, respectively.}: 
\begin{equation}
h=\frac{v}{\sqrt{1-\gamma/3}},\;\;\;\; 
\sqrt{2\gamma}=\epsilon \frac{h^{\prime}}{h},
\end{equation}
and/or the second-order-equation for $y = y(x)$: 
\begin{equation}
\ddot{y}+3h\dot{y}+\frac{3}{2}\epsilon^{2}vv^{\prime}=0.
\end{equation}
Eq.(18) yields the relationship between two functions: 
\begin{equation}
2\dot{y} = - \epsilon^{2} h^{\prime}.
\end{equation}

The inflation condition (5) allows to find the inflationary solution of
eq.(21) via the decomposion over small $\gamma$: 
\begin{equation}
h=v\left(1+\frac 16{\gamma}+o(\gamma)\right),
\end{equation}
\begin{equation}
a=-\frac{1}{H\eta}\left(1+\gamma+o(\gamma)\right),
\end{equation}
where 
\begin{equation}
\sqrt{2\gamma}=\frac{\epsilon v^{\prime}/v}{1-\vartheta/3},\;\;\;\;
\vartheta\equiv\frac{\epsilon\left(\sqrt{\gamma/2} \right)^{\prime}}{1-
\gamma/3}=\frac{\left(\sqrt{2\gamma}\right)^{ \prime}_{\varphi}}{1-\gamma/3}.
\end{equation}
Making use of eqs.(23), (25) we may also present the derivatives of 
$y$-function over the conformal time, 
\begin{equation}
\frac{dy}{d\ln\vert\eta\vert} = \epsilon\sqrt{\frac{\gamma}{2}}
\left(1+\gamma+o(\gamma)\right),\;\;\;\; \vartheta=\frac{d\ln\sqrt{\gamma}}{
d\ln\vert\eta\vert} \left(1-\frac {2}{3}\gamma+o(\gamma )\right).
\end{equation}
We will also need for further analysis the $\varphi$-derivatives at the
horizon-crossing\footnote{%
Eq.(18) yields 
\[
\frac{d\varphi}{d\ln a}=-\sqrt{2\gamma},\;\;\; \frac{d^{2}\varphi}{d(\ln
a)^{2}}=\frac{d\gamma}{d\varphi}, 
\]
the (-) sigh implies that $\varphi$ decreases with time.}, 
\begin{equation}
\frac{d\varphi}{d\ln k}=-\sqrt{2\gamma}\left(1+\gamma+o(\gamma)
\right),\;\;\;\; \frac{d\ln\gamma}{d\ln k}=-2\vartheta\left(1+\frac{2}{3}
\gamma+ o\left(\gamma\right)\right),
\end{equation}
and the scattering potentials (cf.eqs.(52)), 
\[
U\equiv\frac{d^{2}\left(a\sqrt{\gamma}\right)}{a\sqrt{\gamma}d \eta^{2}}%
=a^{2}H^{2}\left(2-\gamma-3\vartheta\left(1-\frac{ \gamma}{3}\right)^{2} + 
\frac{1}{4}\epsilon^{2}\gamma^{\prime\prime}\right), 
\]
\begin{equation}
U^{\lambda}\equiv\frac{d^{2}a}{ad\eta^{2}}=a^{2}H^{2}\left(2- \gamma\right)=
\frac{2}{\eta^{2}}\left(1+\frac {3}{2}\gamma+ o(\gamma)\right).
\end{equation}

Actually, eqs.(24)-(29) are true during the whole period of inflation based
on inequality (5); they describe the evolution along the attractor
inflationary separatisa towards which any solution of eqs.(17)-(19) tends
during the Universe expansion.

However, it is an additional to the inflation condition (5) assumption known
as the slow-roll approximaion, 
\begin{equation}
\vert\vartheta\vert < 1,
\end{equation}
that, when works, simplifys the situation allowing to relate $\gamma$ and $y$
algebraically (see eqs.(26)) and thus to solve eqs.(21), (26) explicitely.
Both inequalities (5) and (30) can be rewritten, respectively, as 
\begin{equation}
\epsilon\frac{v^{\prime}}{v}<1,\;\;\;\;{\rm and}\;\;\;\; 
\epsilon^{2}\frac{\vert v^{\prime\prime}\vert}v<1.
\end{equation}

$\Lambda$-inflation proceeds most difficult near $y\sim 1$. Indeed, for the
power-law potential (15), $v=\sqrt{1+y^{\kappa} }$, the first inequality
(31) meets at the worst point $y\sim y_1 = (\kappa -1)^{\frac{1}{\kappa}}
\simeq 1$ only for small $\epsilon$, 
\begin{equation}
\epsilon<\epsilon_{0}=\frac {2}{\kappa-1},\;\;\;\;or\;\;\;\;
\varphi_{cr}\;{}^>_\sim\;(\kappa-1)\ge 1,
\end{equation}
that we assume hereafter. The second inequality (31) holds at any $y$ unless 
$\kappa<3$. For the latter case the slow-roll approximation is broken in the
field interval 
\begin{equation}
\exp\left(-\frac{1}{\kappa-2}\right)<y<1,
\end{equation}
where the left-hand-side keeps constant: $\frac{\epsilon^{2}v^{ \prime
\prime}}{v}\sim\epsilon^{2}$ (hence, the slow-roll approximation is 
restored in the limit $\epsilon\rightarrow 0$).

So, for $\kappa=2$, the whole evolution for $y<1$ deviates strongly from the
slow-roll approximation. Before coming to it, we write down the evolution
for $\kappa\ge 3$.

\subsection{The $\Lambda\lambda$-Inflation}

The slow-roll approximation is met for $\kappa\ge 3$; then, under conditions
(5) and (30), eq.(23) is integrated explicitely: 
\begin{equation}
a = \exp\left(-\int\frac{d\varphi}{\sqrt {2\gamma}}\right) \simeq 
\gamma^{\frac{1}{6}}\exp\left(-\frac{2}{\epsilon^{2}} \int\frac{vdy}{v^{
\prime}}\right).
\end{equation}
Substituting here $v=\sqrt{1+y^\kappa}$, we have at the horizon-crossing: 
\begin{equation}
\kappa\ge 3:\;\;\;\; y^{2}\left(1-\left(\frac{y_{2}}{y}\right)^{\kappa}
\right)=\Theta,
\end{equation}
where 
$y_{2} = \left(\frac{2}{\kappa-2}\right)^{\frac{1}{\kappa}} \simeq 1$, 
$\Theta = -\frac{\kappa\epsilon^{2}}{2} \ln K = \frac{\kappa-4 }{\kappa-2} - 
\frac{\kappa\epsilon^{2}}{2} \ln K_{c}$, 
$K = \frac{a}{\gamma^{\frac{1}{6}}} = \left(\frac {k}{k_{2}} \right) \left(
\frac{y_2}{y}\right)^{\frac{\kappa-1}{3}} \left( \frac{\kappa/(\kappa-2)}{1+
y^{\kappa}}\right)^{\frac {1}{6}} \sim \frac{k}{k_{2}}$, 
$K_{c} = \left(\frac{k}{k_{cr}}\right) \left(\frac{1}{y}\right)^{\frac{
\kappa-1}{3}} \left(\frac{2}{1+y^{\kappa}} \right)^{\frac {1}{6}} \sim 
\frac{k}{k_{cr}}$. 
Evidently, 
\[
\frac{d\ln K_{(c)}}{d\ln k} = 1+\gamma + \vartheta /3+o(\gamma)
+o(\vartheta)\simeq 1, 
\]
\[
y\simeq\Biggl\{
\begin{array}{lcl}
\Theta^{\frac {1}{2}}, & \; & y>y_{2}, \\ 
\left(\frac{2}{\left(\kappa-2\right)\vert\Theta\vert}\right)^{\frac{1}{
\kappa-2}}, & \; & y<y_{2}.
\end{array}
\]
The transition period between these two asymptotics, $\vert\Theta
\vert^{<}_{\sim}1$, is pretty small in $y$-space, 
\[
\vert y-y_{2}\vert<\frac {1}{\kappa}:\;\;\;\; y\simeq y_{2} + \frac{1}{
\kappa y_{2}}\Theta\simeq 1 - \frac{\epsilon^{2}}{2}\ln K, 
\]
however, it is big in the correspondent frequency band (cf.eq.(32)): 
\begin{equation}
\vert\ln K\vert < \frac{2}{\kappa\epsilon^2} \left({}^{>}_{\sim}\frac{1}{
\epsilon}\right).
\end{equation}

An interesting physical case here is the case with self-interacting field,
which we call $\Lambda\lambda$-inflation: 
\begin{equation}
\kappa=4:\;\;\;\;\;y^{2}\simeq\sqrt{1+(\epsilon^{2}\ln K)^{2}} -
\epsilon^{2}\ln K,
\end{equation}
where $K=K_{c} = \frac{k}{k_{cr} y}(\frac {2}{1+y^{4}})^{\frac{1}{6}}$.
Recall that the $\epsilon$-parameter should not exceed unity if we want to
keep inflation everywhere.

\subsection{The $\Lambda m$-Inflation}

The case of massive field ($\kappa=2$, $v=\sqrt{1+y^2}$) violates the
slow-roll condition and requires more carefull investigation.

The slow-roll approximation works well for $y^{>}_{\sim} 1$, but is broken
at small $y$ as $\frac{v^{\prime\prime}}{v}=v^{-4}\sim 1$ for $y < 1$ (see
eq.(33)). In the latter case $h\simeq 1$ and eq.(22) turns to linear one
presenting the $y$-function as a linear superposition of the {\it fast} (+)
and {\it slow} (-) exponents ($\sim e^{-1.5(1\pm p)x}\sim
\vert\eta\vert^{1.5(1\pm p)}$). This allows for a straightforward, i.e.
independent of the exponent amplitudes, derivation of the $U$-potential at
the dS stage (see eqs.(27),(29)): 
\begin{equation}
y<1:\;\;\;\; U\equiv\frac{d^{2}(a\sqrt{\gamma})}{a\sqrt\gamma d \eta^{2}}
\simeq \frac{d^{3}y}{d\eta^{3}}\left(\frac{dy}{d\eta} \right)^{-1}\simeq
\frac{9p^{2}-1}{4\eta^{2}},
\end{equation}
where $p=\sqrt{1-\frac{2\epsilon^{2}}{3}}$.

The inflationary evolution proceeds in a non-oscilatory way for $\varphi <
\varphi_{cr}$ if 
\begin{equation}
0<p<1,\;\;\;\;\varphi_{cr}\;{}^{>}_{\sim}1.6,
\end{equation}
that we will assume further on. With such a requirement the inflation is
garanteed for any $\varphi$ (cf.eqs.(32)).

To find the exponent amplitudes for $y(<1)$ we have to match the full
inflationary separatrisa at $y\sim 1$. To do it let us exclude the
first-derivative term in eq.(22) introducing a new variable $z = z(\eta)
\equiv ya$: 
\begin{equation}
\frac{d^{2}z}{d^{2}\eta^{2}}-\tilde {U} z=0,
\end{equation}
and then approximate the $\tilde {U}$-function by a simple step-function: 
\[
\tilde {U}\equiv\left(aH\right)^{2}\left(2-\gamma- \frac{3\epsilon^{2}}{
2h^{2}}\right) \simeq \frac{2}{\eta^{2}} \left(1-\frac{3\epsilon^{2}}{4v^{2}}
\right)\simeq\frac{1}{\eta^{ 2}}\Biggl\{
\begin{array}{lcl}
2,   & \; & \eta<\eta_{3}, \\ 
\frac{9p^{2}-1}{4}, & \; & \eta>\eta_{3}, 
\end{array}
\]
where $\eta_{3}\simeq\eta_{cr}$. The solution of eq.(40) is then got
explicitely; matching $z$-function and its first derivative at $%
\eta=\eta_{3} $ and taking into account that $H_{0}z\eta \rightarrow -1$ for
large $y$, we obtain at the dS stage (cf.eqs.(27)): 
\begin{equation}
\omega>1:\;\;\; 
y\simeq\omega^{-\frac{3}{2}}\left({\rm ch}\mu +\frac {1}{p}{\rm sh}\mu
\right),\;\;\; 
\sqrt{\frac{\gamma}{2}}\simeq\frac{\epsilon}{p}\omega^{-\frac{3}{ 2}}
{\rm sh}\mu,\;\;\; 
\vartheta \simeq\frac {3}{2}\left(1-p{\rm cth}\mu\right),
\end{equation}
where $\mu=\frac {3}{2} p\ln\omega^{>}_{\sim} p$, $\omega = \frac{\eta_{3}}{
\eta}\simeq\frac{\eta_{cr}}{\eta}\simeq \frac{k}{k_{cr}}$.

The fitting coefficients in eq.(41) describe a part ($y<1$) of the full
inflationary separatrisa extending from large to small values of the 
$\varphi $-field\footnote{ The fitting accuracy is quite satisfactory. 
Say, in the slow-roll approximation $p \rightarrow 1$: $\frac{\eta_{3}}{
\eta_{cr}}=2^{-\frac {1}{6}}\simeq 1$ and $y\omega^{1.5(1-p)} = \sqrt {e} 
\sim 1$. See the Appendix for more detail.}. We see that at the de Sitter 
stage the function $\vartheta = \vartheta(\omega)>0$ varies slowly, 
\begin{equation}
y<1:\;\;\;\;\;\;\;\vartheta\simeq\Biggl\{ 
\begin{array}{lcl}
\frac {3}{2}-\frac{1}{\ln\omega}, &\; &1^<_\sim \ln\omega<\frac{2}{3p} \\ 
\frac{\epsilon^{2}}{1+p}, & \; & \ln\omega^>_\sim\frac{2}{3p}
\end{array},
\end{equation}
and 
\begin{equation}
\sqrt{2\gamma}=\frac{\epsilon y}{1-\frac{\vartheta}{3}}\;,\;\;\;\;\; 
y\simeq\Biggl\{ 
\begin{array}{lcl}
\frac {3}{2}\omega^{-\frac {3}{2}}\ln\omega, & \; & 1_{\sim}^{<}\ln 
\omega< \frac{2}{3p} \\ 
\frac{1+p}{2p}\omega^{\frac {3}{2}(p-1)}, & \; & \ln\omega\;^{>}_{ \sim}
\frac{2}{3p}
\end{array}.
\end{equation}
The field evolution approaches the slow exponent only for $\ln \omega>
\frac{2}{3p}\;\left(y<\frac{\exp\left(-\frac{1}{p}\right) }{p}\right)$: 
\begin{equation}
y\ll 1:\;\;\;\; 
y\simeq \frac{1+p}{2p}\omega^{\frac {3}{2}(p-1)}, \;\;\;\; 
\sqrt{2\gamma}\simeq \frac{\epsilon}{p}\omega^{\frac{3}{2}(p-1)}.
\end{equation}
For $p\in \left(\frac 23, 1\right)$ the true evolution at the dS stage is 
presented only by the bottom lines in eqs.(42), (43); this fact is used in 
the Appendix to restore the whole inflation dynamics for 
$\epsilon^2<\frac 56$.

Comparing eqs.(35) and (44) we see that at the dS stage $y$ decays as $\ln k$
for $\kappa\ge 3$, whereas it is the power-law for $\kappa=2$. For the
intermediate case $2<\kappa <3$ the slow-roll approximation is violated only
within the limited interval (33) where the solution can be matched by
eq.(41) with $p=\sqrt{1 - \frac{\kappa\epsilon^{2}}{3}}$.

\section{The generation of primordial perturbations}

Below, we introduce the S and T metric perturbation spectra and find them
for $\Lambda$-inflation.

The linear perturbations over the geometry (16) can be irreducably
represented in terms of the uncoupled Scalar, Vector and Tensor parts [1].
The vector perturbations are not induced in our case as scalar fields are
not their sources. Under the action (3) we are rest with only the S and T
modes, and the new geometry looks as follows: 
\begin{equation}
ds^{2} = (1+h_{00})\;dt^{2}+2ah_{0\alpha}\;dtdx^{\alpha}-a^{2}
(\delta_{\alpha\beta} + h_{\alpha\beta})\;dx^{\alpha}dx^{\beta},
\end{equation}
\[
\frac {1}{2}h_{\alpha\beta}=A\delta_{\alpha\beta}+B_{,\alpha
\beta}+G_{\alpha\beta},\;\;\;\; h_{0\alpha}=C_{,\alpha}, 
\]
where $G^{\alpha}_{\alpha}=G^{\beta}_{\alpha,\beta}=0$. The gravitational
potentials $h_{00}$, $A$, $B$, $C$ are coupled to the perturbation of scalar
field $\delta\varphi$, whereas $G_{ \alpha \beta}$ is the free tensor field.
The Lagrangian $L^{(2)}$ of the perturbation sector of the geometry (45) is
given by decomposing the integrand (3) up to the second order in the
perturbation amplitudes. Our further analysis of the S-sector follows a
general theory of the $q$-field ([4], [16]), the gravitational waves are
totally described by the gauge-independent 3D-tensor $G_{\alpha\beta}$ ([3],
[17], [18]).

Instead of considering gauge-dependent potentials ($h_{00}$, $A$, $B$, $C$, 
$\delta\varphi$) we introduce the gauge-invariant canonical 4D-scalar $q$
uniquely fixed by the appearence of the S-part of the perturbative
Lagrangian $L^{(2)}$ similar to a massless field: 
\begin{equation}
L^{(2)} = L(q,G_{\alpha\beta}) = \frac {1}{2}\alpha^{2}q_{,i} q^{,i}+
\frac{1}{2}G_{\alpha\beta,\gamma}G^{\alpha\beta,\gamma},
\end{equation}
where 
$\alpha^{2}\equiv 2\gamma = \frac{\rho+p}{H^{2}} = \left(\frac{\dot{\varphi
}}{H}\right)^{2}$, 
$\alpha = \frac{\dot{\varphi}}{H}$ (mind the choice of the sign for 
$\alpha$ that we take coinciding with the sign of $\dot{\varphi}$). The
relation of $q$ to the original potentials takes the following form: 
\[
\delta\varphi=\alpha\left(q+A\right),\;\;\;\;
a^{2}\dot{B}+C= \frac{\Phi+A}{H}, 
\]
\begin{equation}
\frac{1}{2}h_{00} = \gamma q + \left(\frac{A}{H}\right)^{.}, \;\;\;\;
\Phi=\frac{H}{a}\int a\gamma q dt,
\end{equation}
\[
\frac{\delta\rho}{\rho+p} = \frac{\dot{q}}{H} - 3(q+A),\;\;\;\; 4\pi
G\delta\rho_{c} \equiv\gamma H\dot{q}=a^{-2}\triangle\Phi, 
\]
where $a$, $\varphi$, $H$, $\alpha$, $\gamma$, $\rho=\frac{1}{2} w^{2}+V$
and $p=\frac{1}{2}w^{2}-V$ are the background functions of time, $\Phi$ is
the "Newtonian" gauge-invariant gravitational potential related non-locally
to $q$, $\triangle\equiv\partial^{ 2}/\partial^{2}\vec{x}^{2}$ is spatial
Laplacian, ($\triangle=- k^{2}$ in the Fourrier representation, 
$\delta\rho_{c}$ is the comoving density perturbation). Any two potential
taken from the triple $A$, $B$, $C$ are arbitrary functions of all
coordinates, which determines the gauge choice. All information on the
physical scalar perturbations is contained in the $q=q(t,\vec{x}) $ field,
the dynamical 4D-scalar propagating in the unperturbed Friedmann geometry
(i.e. independently of any gauge in eq.(45)).

The equations of motions of the $q$ and $G_{\alpha\beta}$ fields are two
harmonic oscilators: 
\begin{equation}
\ddot{q}+\left(3H+\frac{\dot{\gamma}}{\gamma}\right)\dot{q}-a^{- 2}\triangle
q=0,
\end{equation}
\begin{equation}
\ddot{G_{\alpha\beta}}+3H\dot{G_{\alpha\beta}}-a^{-2}\triangle
G_{\alpha\beta}=0.
\end{equation}
A standard procedure to find the amplitudes generated is to perform the
secondary quantization of the field operators, 
\begin{equation}
q =\int^{\infty}_{-\infty}d^{3}\vec {k}\left(a_{\vec k}q_{\vec k}+
a^{+}_{\vec{k}}q^{\ast}_{\vec k}\right),
\end{equation}
\[
G_{\alpha\beta}=\sum_{\lambda}\int^{\infty}_{-\infty}d^{3}\vec{k}
\left(a^{\lambda}_{\vec k}h^{\lambda}_{\vec{k}\alpha\beta}+
a^{ \lambda +}_{\vec{k}}h^{\lambda \ast}_{\vec {k}\alpha\beta}\right
), 
\]
where $+/\ast$ denotes Hermit/complex conjugation, index $\lambda =+,\times$
runs two polarizations of gravitational waves with the polarization tensors 
$c_{\alpha \beta}^{\lambda}(\vec{k})$, and 
\begin{equation}
q_{\vec{k}}=\frac{\nu_{k}}{\left(2\pi\right)^{\frac{3}{2}}\alpha a}\;
e^{i\vec {k}\vec {x}},
\end{equation}
\[
h^{\lambda}_{\vec{k}\alpha\beta} = 
\frac{\nu_{k}^{\lambda}}{\left(2\pi\right)^{\frac{3}{2}}a}\; 
e^{i\vec{k}\vec{x}}c^\lambda_{\alpha\beta}\left(\vec{k}\right), 
\]
\[
\delta^{\alpha\beta}c^\lambda_{\alpha\beta}\left(\vec k\right)= k^\alpha
c^\lambda_{\alpha\beta}\left(\vec k\right)=0, \;\;\;\;
c_{\alpha\beta}^{\lambda}(\vec{k})c^{\alpha\beta\lambda^{\prime}} 
\left(\vec{k}\right)^\ast = \delta_{\lambda\lambda^{\prime}}. 
\]
The time-dependent $\nu$-functions satisfy the respective Klein-Gordon
equations, 
\begin{equation}
\frac{d^{2}\nu_{k}^{(\lambda)}}{d\eta^{2}}+\left(k^{2}-U^{(
\lambda)}\right)\nu^{(\lambda)}_{k}=0,
\end{equation}
with $U=U(\eta)\equiv\frac{d^{2}\left(\alpha a\right)}{\alpha ad\eta^{2}}$
for the $q$-field and $U^{\lambda} = U^{\lambda} (\eta)\equiv\frac{d^{2} 
a}{ad\eta^{2}}$ for each polarization of the gravitational waves 
$\nu^{\lambda}_{k}$. The standard commutation relations between the
annihilation and creation operators, 
\[
\left[ a_{\vec{k}}a^{+}_{\vec{k}^{\prime}}\right]=\delta\left( \vec{k}-
\vec{k}^{\prime}\right),\;\;\;\; \left[ a^{\lambda}_{\vec {k}}
a^{\lambda^{\prime}+}_{\vec {k}^{ \prime}}\right] = \delta\left(\vec{k} - 
\vec{k}^{\prime}\right) \delta_{\lambda \lambda^{\prime}}, 
\]
require the following normalization condition for each of the 
$\nu$-functions: 
\[
\nu_{k}^{(\lambda)}\frac{d\nu^{(\lambda)\ast}_{k}}{d\eta}-
\nu_{k}^{(\lambda)\ast}\frac{d\nu_{k}^{(\lambda)}}{d\eta}= i. 
\]

Eqs.(46)-(52) specify the {\it parametric amplification effect}: the
production of the perturbations -- the phonons for $S$-mode [4] and the
gravitons for $T$-mode [3] -- in the process of the Universe expansion (the
latter is imprinted in the non-zero scatering potentials $U^{(\lambda)}$ in
eqs.(52)).

From the inflationary condition (5) one finds always $k\eta \rightarrow
-\infty$ for the early inflation (scales inside the horizon); therefore, the
microscopic vacua states of the $q$ and $G_{\alpha\beta}$ fields mean the
positive frequency choice for the initial $\nu$-functions: 
\begin{equation}
k\vert\eta\vert\gg 1:\;\;\;\; 
\nu_{k}^{(\lambda)} = \frac{\exp(-ik\eta)}{\sqrt{2k}}.
\end{equation}
So, the problem of the spontaneus creation of density perturbations and
gravitational waves is finally reduced to solving the eqs.(52), (53) with
the effective potentials $U^{(\lambda)}$ taken from the inflationary
background regimes considered above.

For the late inflation $k\eta\rightarrow 0$ (scales outside the horizon),
the perturbations become semiclassical since the fields are getting frozen
in time and thus acquire the phase (only the ${\it {growing}}$ solutions of
eqs.(48),(49) survive in time)\footnote{ Here, the transfer from the 
quantum (squeezed) to classical case occurs when one neglects the 
{\it decaying} solutions of eqs.(48),(49) for $\eta\rightarrow 0$: 
\[
k\vert\eta\vert < 1:\;\;\;\; q_{d}\sim\int^{0}\frac{d\eta}{a^{2}\gamma} = 
\frac{1}{3\gamma} H^{2}\eta^{3}\left(1 + O(\gamma)\right),\;\;\;\;
G_d\sim\int^{0}\frac{d\eta}{a^{2}} = \frac {1}{3}H^{2}\eta^{3}
\left(1+O(\gamma)\right), 
\]
and thus is left only with the growing ones (see eq.(54)). This procedure
turns the annihilation and creation operators into the $C$-numbers (where
the commutators vanish).}, 
\begin{equation}
k\vert\eta\vert\ll 1:\;\;\;\; q=q(\vec {x}),\;\;\;\; G_{\alpha\beta} =
G_{\alpha\beta}(\vec {x}).
\end{equation}
One can, therefore, treat these time-independent perturbation fields as
realizations of the classical random Gaussian fields with the following
power spectra: 
\[
\langle q^{2}\rangle = \int^{\infty}_{0}q^{2}_{k}\frac{dk}{k}, \;\;\;\;
\langle G_{\alpha\beta}G^{\alpha\beta}\rangle=\int^{\infty}_{0} h^{2}_{k}
\frac{dk}{k}, 
\]
\begin{equation}
q_{k}=\frac{k^{\frac{3}{2}}\vert\nu_{k}\vert}{2\pi a\sqrt{\gamma }},\;\;\;\;
h_{k}=\frac{k^{\frac{3}{2}}\sqrt{\vert\nu^{+}_{k}\vert^{2}+\vert
\nu_{k}^{\times}\vert^{2}}}{\pi a\sqrt{2}}=\frac{k^{\frac{3}{2}}
\vert\nu_{k}^{\lambda}\vert}{\pi a}
\end{equation}
Here the $\nu$-functions are taken in the limit $\vert\eta\vert\ll k^{-1}$,
and the gravitational wave spectra in both polarizations are identical. The
local slopes and the ratio of the power spectra are found as follows: 
\begin{equation}
n_{S}-1 \equiv 2\frac{d\ln q_{k}}{d\ln k},\;\;\;\; n_{T} \equiv 2\frac{d\ln
h_{k}}{d\ln k}, \;\;\;\; r \equiv \left(\frac{h_{k}}{q_{k}}\right)^{2} =
4\left(\gamma\vert \frac{\nu_{k}^{\lambda}}{\nu_{k}}\vert^{2}\right)_{k\vert
\eta\vert \ll 1}.
\end{equation}

Note, that the quantities $q_k$, $h_k$, $n_S$, $n_T$, $r$ are the functions
of the wavenumber only. For references, we recall also the density
perturbation and Newtonian potential linked to the $q$-field in the
Friedmann Universe (cf. eqs.(47), (54)), 
\[
k<aH:\;\;\;\;\;\;\;\; \Delta_k=\frac{2}{3}\left(\frac{k}{aH}%
\right)^2\Phi_k,\;\;\;\; \Phi_k=\Gamma q_k, 
\]
where $\Delta_k$, $\Phi_k$ are the dimentionless spectra, respectively, 
\[
\left\langle\left(\frac{\delta\rho_c}{\rho}\right)^2\right\rangle =
\int_0^\infty\Delta_k^2\frac{dk}{k},\;\;\;\; \langle \Phi^2\rangle=
\int_0^\infty\Phi_k^2\frac{dk}{k}, 
\]
and $\Gamma=\frac{H}{a}\int a\gamma dt = 1-\frac{H}{a}\int a\,dt$ is the
function of time ($\Gamma=(1+\beta)^{-1}=$ const for the power-law
expansion, $a\sim t^\beta$).

\section{The power spectra}

When it works, the slow-roll approximation allows for simple derivation of
the S-spectrum ($U^{(\lambda)}\simeq 2/\eta^2$, cf.eqs.(29)): 
\begin{equation}
q_{k}\simeq\frac{H}{2\pi\sqrt{2\gamma}},\;\;\;\; 
h_{k}=\frac{H}{\pi\sqrt{2}},\;\;\;\;k=aH,
\end{equation}
where $H=H_{0}v$, $\sqrt{2\gamma}\simeq\epsilon\frac{v^{\prime}}{ v}$, 
$\vartheta\simeq\frac{1}{2}\epsilon^{2}\left(\frac{v^{ \prime}}{v}\right)^{
\prime}$. The spectra ratio and the local slopes are then the
following (see eqs.(28), (32), (56)): 
\begin{equation}
r\simeq -2n_{T} = 4\gamma\simeq \frac{1}{2}\left(\frac{\epsilon \kappa
y^{\kappa-1}}{1+y^{\kappa}}\right)^{2} \le r_{max}= 
\frac{1}{2}\left(\frac{\epsilon\left(\kappa-1\right)}{y_{1}} \right)^2
\simeq 2\left(\frac{\epsilon}{\epsilon_0}\right)^2,
\end{equation}
\[
n_S-1\simeq 2\left(\vartheta-\gamma\right)=f\left(y\right), \;\;\;\;
f_{-}\le f\left(y\right)\le f_{+}, 
\]
where 
$f\left(y\right) = \frac{\kappa}{2}y^{\kappa-2}\left(\frac{\epsilon }{1+
y^{\kappa}}\right)^2\left(\kappa-1-\frac{\kappa+2}{2}y^\kappa \right)$, 
$y_{\pm}=\left(\kappa-1\mp\kappa\sqrt{\frac{\kappa-1}{\kappa+2}} 
\right)^{\frac{1}{\kappa}}$, 
$f_{\pm}=f\left(y_{\pm}\right)=\frac{\left(\kappa-1\right)\left( \kappa+2
\right)}{12}\left(\frac{\epsilon}{y_{\pm}}\right)^2\left( \pm 2\sqrt{\frac{
\kappa-1}{\kappa+2}}-1\right)\simeq\pm\left( \frac{\epsilon}{\epsilon_0}
\right)^2$.

Eqs.(58) are true for $v=\sqrt{1+y^{\kappa}}$; the T-spectrum deviates
maximally from HZ (and the spectrum ratio reaches its maximum) at 
$y_{1}\simeq (\kappa-1)^{\frac{1}{\kappa}}\simeq 1$; the S-spectrum achieves
its minimum and becomes exactly HZ one at 
$y_{4} = \left(\frac{\kappa-1}{1+\frac{\kappa}{2}}\right)^{\frac{ 1}{
\kappa}}= y_{1}(\frac{2}{\kappa+2})^{\frac {1}{\kappa}}\simeq 1$, 
it is the most red (blue) at $y_{-}$ ($y_{+}$); the points $y_1$ and 
$y_4$ lay always inside the interval $\left(y_{+},y_{-}\right)$ while 
the region (36) resides there only if $\kappa\le 8$. Eq. 
$f\left(y\right)=$ const $\in\left[f_{-},f_{+}\right]$ has two solutions: 
one is located within the interval $\left[y_{+},y_{-}\right]$ where $r$ is
large, $\frac{r}{r_{max}}^>_\sim\left(\frac{\kappa+1}{3\kappa}\right)^2$ and 
$r(n_S=1)\simeq r_{max}$; another is outside this interval where $r$ is
small, $\frac{r}{r_{max}}<1$ and $r(n_S=1)=0$.

So, for $\kappa\ge 3$ we have from eq.(35) the following asymptotics for the
power spectra: 
\[
q_{k}^{2}\simeq\left( \frac{H_{0}}{\epsilon\pi\kappa}\right)^{2} \frac{
\left(1+y^{\kappa}\right)^{3}}{y^{2\kappa-2}}\simeq\frac{ \lambda_{\kappa}}{
12\pi^{2}}\Biggl\{
\begin{array}{lcl}
\kappa^{\frac{\kappa-4}{2}}\vert 2\ln K\vert^{\frac{\kappa+2}{2}}, & \; & 
K<\exp \left(-\frac{2}{\kappa\epsilon^{2}}\right) \\ 
\left(\frac{V_{0}}{\lambda_{\kappa}}\right)^{\frac{\kappa-4}{ \kappa-2}
}\left(\left(\kappa-2\right)\ln K\right)^{\frac{2\kappa -2}{\kappa-2}}, & \;
& K>\exp\left(\frac{2}{\kappa\epsilon^{2}} \right)
\end{array}
, 
\]
\[
h_{k}^{2}=\frac{H_{0}^{2}}{2\pi^{2}}\left(1+y^{\kappa}\right) = 
\frac{1}{6\pi^{2}}\Biggl\{
\begin{array}{lcl}
\frac{\lambda_{\kappa}}{\kappa}\vert 2\kappa\ln K\vert^{\frac{ \kappa}{2}},
& \; & K<\exp\left(-\frac{2}{\kappa\epsilon^{2}} \right) \\ 
V_{0}, & \; & K>\exp\left(\frac{2}{\kappa\epsilon^{2}}\right)
\end{array}
. 
\]
In the transition region (36) the ratio of the spectra is approximately
constant independent of the $\kappa$-index: $r\simeq 2\epsilon^{2}$ (it is a
factor $\epsilon^{-2}_0$ less than $r_{max}$).

For $\Lambda\lambda$-inflation the spectra are resolved explicitely (see
eq.(37)): 
\begin{equation}
\kappa=4:\;\;\;\; 
\begin{array}{l}
q_{k}\simeq\frac{1}{\pi}\sqrt{\frac{2\lambda}{3}}\left(
\epsilon^{-4}+\ln^{2}K\right)^{\frac{3}{4}}, \\ 
h_{k}=\frac{H_{0}}{\pi}\left(1+\frac{\ln K}{\sqrt{\epsilon^{-4} +\ln^{2}K}}
\right)^{-\frac{1}{2}}, 
\end{array}
\end{equation}
and 
$y_{1}=3^{\frac 14}$, 
$K_1=\exp\left(-\frac{1}{\sqrt{3} \epsilon^2}\right)$, 
$y_{2}=y_{4}=1$, 
$y_{-}=\frac{1}{y_{+}}= \left(\sqrt{2}+1\right)^{\frac{1}{2}}$, 
$K_{\pm}=\exp\left(\mp \frac{1}{\epsilon^2}\right)$. 
An example of the power spectra for $\epsilon = 0.3$ is shown in Fig.1. 
Fig.2 clarifys the relation between $r$ and $n_S-1$ for any $\epsilon<1$. 
We see there is no correlation between the blueness and large $r$: the 
region of large $r$-values is located in the red and HZ sectors of 
the S-spectrum.

\begin{figure}
\epsfxsize=13cm
\centerline{\epsfbox{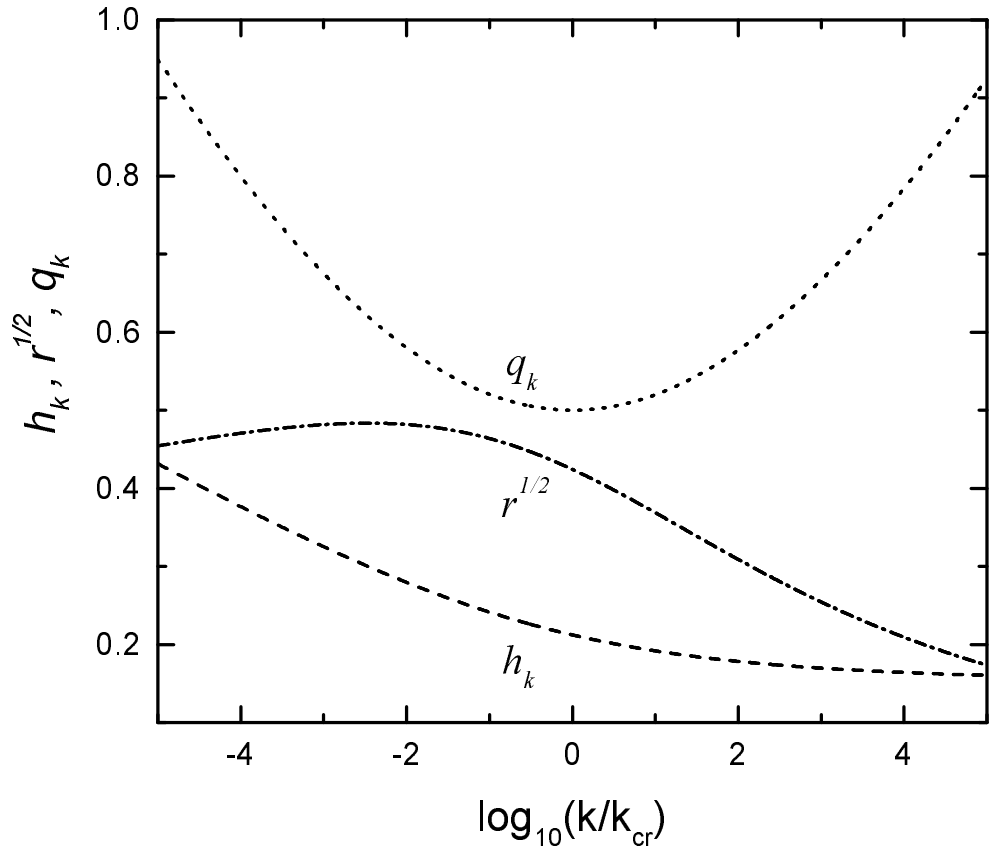}}
\caption{The spectra of scalar perturbations $q_{k}$ (dotted 
curve), tensor perturbations $h_{k}$ (dashed curve), and the 
ratio between them $r^{\frac {1}{2}}\equiv h_{k}/q_{k}$ (dot-dashed 
curve), in the $\Lambda\lambda$-inflation model with $\epsilon=0.3$. 
The normalization is arbitrary, however the ratio does not depend 
on normalization and is true for the used model parameter.}
\end{figure}

\begin{figure}
\epsfxsize=13cm
\centerline{\epsfbox{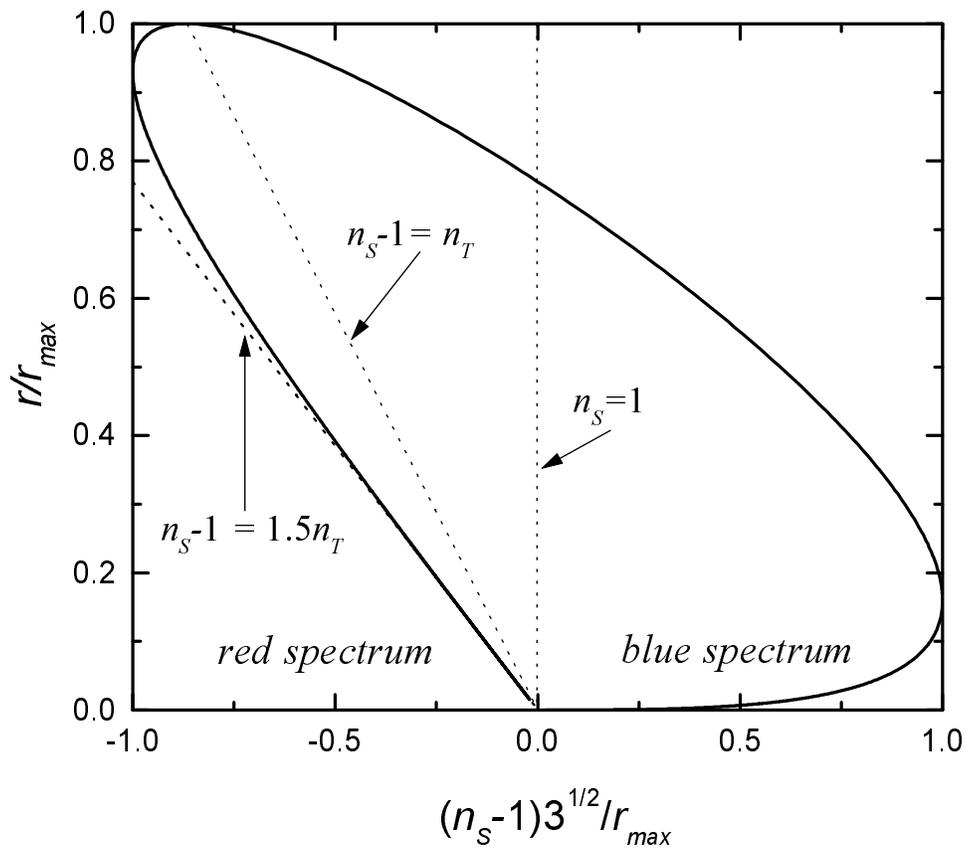}}
\caption{The relationship between $r$ and $n_S$ for $\Lambda$-inflation
($r_{max}=\frac{3\sqrt 3}2 \epsilon^2$, $r=-2n_T$).}
\end{figure}

Let us now turn to the case where the slow-roll approximation is broken.

For $\Lambda m$-inflation eqs.(57) are true except the blue part of the
S-spectrum ($k>k_{cr}$) where it must be corrected. Here eqs.(52), (53) are
solved explicitely, 
\[
y<1:\;\;\;\; 
k^{\frac {3}{2}}\nu_{k}\simeq\frac{ik\sqrt{\pi x} }{2} H^{(1)}_{\frac{3}{2}p}
(x)\;\;
{}^{\longrightarrow}_{x\ll 1} \;\;
\frac{caH_{0}}{\sqrt{2}p}x^{\frac {3}{2}(1-p)}, 
\]
where $H^{(1)}_p(x)$ is the Hankel function, $x=k\vert\eta\vert$, 
$c=\frac{p}{\sqrt{2\pi}}\Gamma\left(\frac {3}{2}p\right)2^{\frac{3 }{2}p} 
=\frac{2^{3p/2}}{3\sqrt{\pi/2}}\Gamma(1+\frac{3}{2}p)$. Taking into account 
the field asymptotics for $y\ll 1$ (see eq.(44)) we obtain the following
S-spectrum in the blue range: 
\begin{equation}
k>k_{cr}:\;\;\;\; q_{k}\simeq\frac{cH_{0}}{2\pi\epsilon}\left(\frac{k}{k_{cr}
} \right)^{\frac 32(1-p)},\;\;\;\;n_S^{blue}=4-3p>1.
\end{equation}
As we see, the spectrum amplitude remains a finite number for $p \rightarrow
0$ ($n_S\rightarrow 4$).\footnote{This corrects the wrong statement on the 
divergence of $q_k$ at $p\rightarrow 0$ made in some previous publications.} 
In most applications we usually have $n_S <3\;\;(p>\frac 13)$; in this case 
the whole spectrum approximation for the $\Lambda m$-inflation looks as 
follows: 
\begin{equation}
q_k=\frac{H_0(1+y^2)^{\frac 12}(\tilde c+y^2)}{2\pi\epsilon y},
\end{equation}
where $\tilde c = \frac{c(1+p)}{p} = \frac{1+p}{\sqrt\pi}\Gamma \left(
\frac {3}{2}p\right)2^{\frac 32(p-1)}$ and $y$ is taken at horizon crossing 
(see eq.(41)).

\section{The T/S effect in $\Lambda$-inflation}

A large T/S $\sim 1$ (when $k_{cr}\in (10^{-4}, 10^{-3})$) is an intrinsic
property of $\Lambda$-inflation. Below, we demonstrate it straightforwardly
for COBE angular scale (see [19], [20], eq.(1)).

Then the S and T are written as follows: 
\begin{equation}
{\rm {S} = \sum_{\ell=2}^{\infty} S_{\ell} \exp\left[- \left(\frac{2\ell + 
1}{27}\right)^2\right],\;\;\; {T} = \sum_{\ell=2}^{\infty} T_{\ell}
\exp\left[- \left(\frac{2\ell + 1}{27}\right)^2\right],}
\end{equation}
where $S_{\ell}$, $T_{\ell}$ are the corresponding variances in $\ell$th
harmonic component of $\Delta T/T$ on the celestial sphere, 
\begin{equation}
S_{\ell} = \sum_{m=-\ell}^{\ell} \vert a_{\ell m}^{(S)} \vert^2, \;\;\;\;
T_{\ell}=\sum_{m=-\ell}^{\ell}\vert a_{\ell m}^{(T)}\vert^2, \;\;\;\; 
\frac{\Delta T}{T}\left(\vec e\right)= 
\sum_{\ell,m,S,T}a_{\ell m}^{(S,T)}Y_{\ell m}\left(\vec e\right).
\end{equation}

The calculations can be done for the instantaneous recombination, 
$\eta=\eta_E$ [2], 
\[
\frac{\Delta T}{T}\left(\vec e\right) = \left(\frac 14\delta_\gamma -
\vec {e}\vec {v}_b+\frac {1}{2} h_{00} \right)_E +\frac12 \int^0_E
\frac{\partial h_{ik}}{\partial\eta}e^ie^kdx, \;\;
e^i =(1, -\vec e),\; x\equiv \vert\vec{x}\vert=\eta_0-\eta, 
\]
where the SW-integral makes a dominant contribution on large scale (see
eq.(45)), $\delta_\gamma$ and $\vec v_b$ are the photon density contrast and
baryon perculiar velocity, respectively. The mean $S_\ell$ and $T_\ell$
values seen by an arbitrary observer in the matter dominated Universe (e.g.
[21]) are explicitely related with the respective power spectra (see
eqs.(55)): 
\begin{equation}
S_\ell = \frac{2\ell+1}{25}\int^{\infty}_0q_k^2 j^2_\ell\left(\frac{k}{k_0}
\right)\frac{dk}{k},
\end{equation}
\begin{equation}
T_\ell =\frac{9\pi^2}{16}\left(2\ell+1\right) \frac{\left(\ell+ 2\right)!}{
\left(\ell-2\right)!}\int^{\infty}_0h_k^2 I^2_\ell\left(\frac{k}{k_0}\right)
\frac{dk}{k},
\end{equation}
where $k_0=\eta_0^{-1}=\frac{H_0}{2}\simeq 1.6\times 10^{-4}h$ Mpc ${}^{-1}$, 
\[
j_\ell\left(x\right)=\sqrt{\frac{\pi}{2x}}J_{\ell+1/2}\left(x
\right),\;\;\;\; I_\ell (x) = \int_0^x\frac{J_{\ell+1/2}\left(x-y\right)}{
\left( x-y\right)^{5/2}}\frac{J_{5/2}\left(y\right)}{y^{3/2}}dy. 
\]

We have derived T/S for $\Lambda m$-inflation using the appoximation (61).
The result is presented in Fig.3 as a function of two parameters of the
model: the spectrum index in the blue asymptotics $n_S^{blue}$ (see eq.(60))
and the critical scale $k_{cr}$ (in units $k_0$). A similar behaviour of T/S
is met for $\Lambda\lambda$-inflation.

\begin{figure}
\epsfxsize=13cm
\centerline{\epsfbox{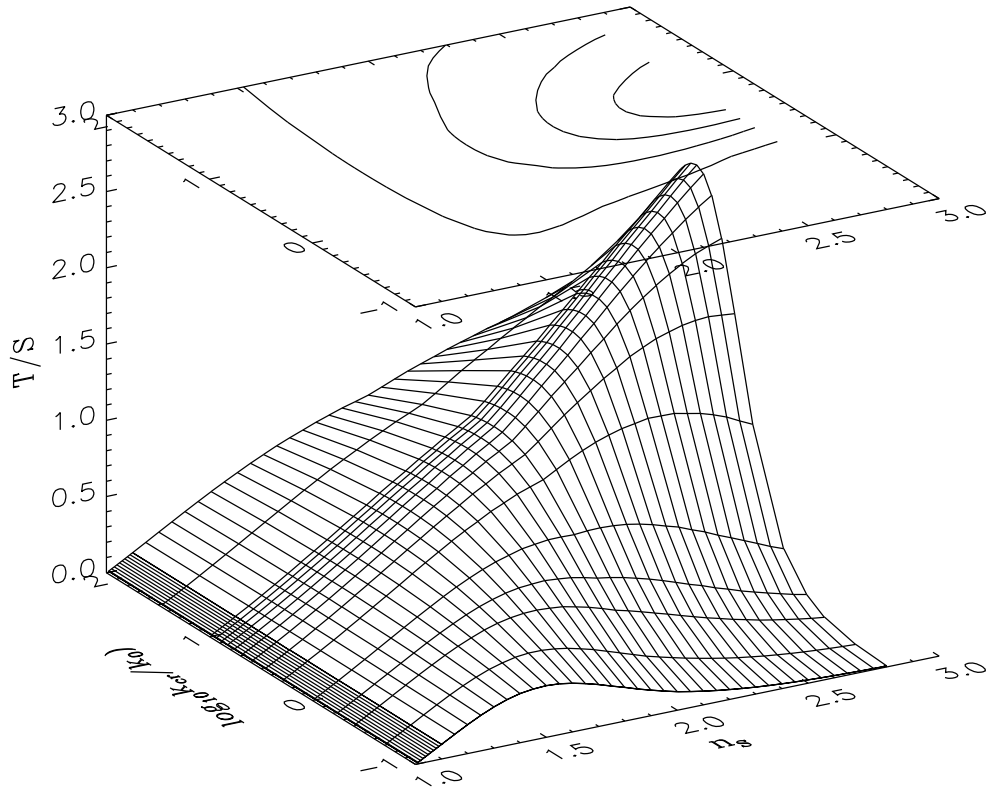}}
\caption{T/S as a function of $k_{cr}$ and $n_S^{blue}$ ($n_S$ in blue 
asymptotic) in the case of $\Lambda m$-inflation.}
\end{figure}

Actually, the two-arm structure of T/S is typical for any $\Lambda$
-inflation model: T/S gets its maximum at $k_{cr}\sim k_{ COBE}\sim 10^{-3}h$
Mpc${}^{-1}$ and gradually decays in both, blue ($k_{cr}<k_{COBE}$) and far
red ($k_{cr}\gg k_{COBE}$) sectors of the S-mode. To be precise, the
T/S-maximum is achieved in the location of $r$-maximum (where $\gamma$ is the
largest and thus $\vartheta=0$). There (and anywhere) the S-spectrum slope
is pretty close to HZ for $\epsilon\ll \epsilon_0$ (cf.eqs.(32), (58)): 
\begin{equation}
1-n_S^{(r_{max})}\simeq -n_T^{(r_{max})}=2\gamma_{max}\simeq \frac {1}{2}
r_{max}\simeq \left(\frac{\epsilon}{\epsilon_0}\right)^2 \ll 1.
\end{equation}
It is important that T/S remains large in a broad $k$-region including the
point where the S-spectrum is exactly HZ: $\frac{r_{n_S = 1}}{r_{max}}\simeq
\frac{4}{9}\left(1+ \frac{\kappa}{2}\right)^{\frac{2}{\kappa}} >\frac 49$.

\section{Discussion}

It seems as a paradox that T/S can be as large as 1 for such a simple model
as $\Lambda$-inflation. However, it can be easily understood. In fact, the
model recalls a case of double inflation where the large T/S is generated in
the intermediate scales between the first and second stages. So, we can
assume that it is sufficient to evaluate T/S by the end of the first stage 
($\varphi \sim\varphi_{cr}\sim 1$) where the slow-roll condition is marginally
applicable. Here (cf.eqs.(2), (32)) 
\begin{equation}
\frac TS \sim \varphi^{-2}\sim 1.
\end{equation}

Often T/S is presented as a function of the gravitational-wave-spectrum
index $n_T$ or the inflationary $\gamma$-parameter estimated in the given
scale, see eq.(2) (e.g. [22], [23], [24], [25], and others). We think this
formula is universal for most types of cosmic inflation. We can argue it by
plainly noting a clear physical equation, 
\begin{equation}
\frac TS\simeq 3r,
\end{equation}
where $r$ is taken in scale where the T/S is determined ($k_{COBE} \sim
10^{-3}h$ Mpc ${}^{-1}$). The factor 3\footnote{%
Or somewhat about 3, to be found more accurately by special investigation
elsewhere. Eventually, it is proportional to the ratio of the effective
numbers of T and S spin projections on given spherical harmonics, see
eqs.(64), (65).} takes into account a higher ability of T-mode to contribute
to $\Delta T/T$. We now see from eqs.(52)-(56) that $r$ is a number found in
the limit $k\vert\eta \vert\ll 1$: 
\begin{equation}
k\vert\eta\vert\ll 1:\;\;\;\; 
r=4\gamma \vert\frac{\nu_k^\lambda}{\nu_{k}}\vert^{2}.
\end{equation}
Implying that the r.h.s. stays frozen outside the horizon ($k\vert \eta\vert
< 1$) we can estimate $r$ as the r.h.s. of eq.(69) at inflation horizon
crossing time. Thus, we may conclude that 
\begin{equation}
r\simeq 4\left(\gamma \vert\frac{\nu_k^\lambda}{\nu_k}\vert^2
\right)_{k\vert\eta\vert=1}\simeq 4\gamma{}_{_{k\vert\eta\vert =1}}.
\end{equation}
The latter is due to the fact that the functions $\nu_k^\lambda$ and $\nu_k$
are close to each other at the horison crossing: they both start from the
same initial conditions (53) and obey the same equations inside the horizon
(see eqs.(52))\footnote{
The difference in their evolutions originates only because of various
effective potentials $U^{(\lambda)}$ entering eqs.(52); however both
potentials vanish for $k\vert\eta\vert>1$.}. Notice this argument is more
general than the slow-roll-condition validity: actually, according to
eqs.(51), (70) the $r$-number counts just the difference between the phase
space volumes of phonons and gravitons.

So, we see that large T/S is created each time when the $\gamma$ factor
approaches subunity values. It may happen either in the end of inflation
(note inflation stops for $\gamma=1$) or in a numerous intermediate periods
during inflationary regime where one type of the inflation is changed for
another one. Such a transition periods can be caused by many reasons; e.g.
due to a functional change of the dynamical potential in the course of
inflation (e.g. $\Lambda$-inflation), or a percularity in the potential
energy shape (e.g. non-analiticity, a step, plateau, or a break of the first
or second derivative of $V(\varphi)$), see e.g. [26]), or a change of the
inflaton field (e.g. double-inflation), or any type of phase transitions or
other evolutionary restructurings of the field Lagrangian that may slower
down, terminate, or break up the process of inflation.

Obviously, each particular way of inflation leaves its own imprints in the
power spectra and requires special investigation. However, the issue of T/S
is a matter of the very generic argument: the inflationary ($\gamma$, $H$)
and/or spectral ($r$, $n_T$) parameters estimated in the appropriate
energy/scale region. It (the T/S value) is totally independent of the local 
$n_S$ and, thus, has nothing to do with a particular S-spectrum shape
produced in a given model.

The principal quantity for estimating T/S becomes the energy of inflaton:
the Hubble parameter at the inflationary horizon-crossing time, $H$ $[GeV]$.
The motivation is the following: as the CGW amplitude is always about $H$
(cf.eq.(57)) and $q_k\sim10^{-5}$ (from LSS originated from the
adiabatic S-mode), we have 
\begin{equation}
\frac{{\rm T}}{{\rm S}}\simeq\frac 16\left(\frac{H}{q_{COBE}}\right)^2
=\left(\frac{H}{6\times 10^{13}GeV}\right)^2\left(\frac{10^{-5}}{ q_{COBE}}
\right)^2,
\end{equation}
where $q_{COBE}\equiv q_{k_{COBE}}$. So, measuring the T/S brings a vital
direct information on the physical energy scale where the cosmic
perturbations has been created; a cosmologically noticable T/S could be
achieved only if the inflation occured at subPlanckian (GUT) energies, $H{}>
10^{13}GeV$. If the CDPs were generated at smaller energies (e.g. during
electroweek transition) then T/S would vanish.

The point we emphasize in this paper is that for $\Lambda $-inflation. It
brings about two distinguished signatures -- a wing-like S-spectrum and the
possibility for large T/S -- under quite a simple and natural assumption on
the potential energy of inflaton: the existence in $V(\varphi )$ of a 
{\it metastabe dynamical constant} in addition to an {\it arbitrary 
functional} $\varphi $-dependent term. It is obviously three independent 
parameters that determine the degrees of freedom of any $\Lambda $-inflation 
model. They can be, for instance, T/S and the local $n_S$ (at the COBE scale)
as well as $k_{cr}$ (the scale where S-spectrum is at minimum) or, 
alternatively, the $r$-maximum and its position (the $k_1$ scale) as well as 
$V_0$. In case if T/S is large, we find quite a defenite prediction on the 
location of the $\Lambda $-inflation parameters near GUT energies 
(see eq.(15)): 
\[
\frac{{\rm T}}{{\rm S}}>0.1:\;\;\;\;
\begin{array}{rcl}
\sqrt{V_0} & \in  & \left( \frac{\zeta ^{-\frac \kappa 2}}{\sqrt{\kappa/2}},
\zeta \right) \left( \frac{q_{COBE}}{10^{-5}}\right) \left( 7\times
10^{15}GeV\right) ^2 \\ 
\frac{\sqrt{\lambda _\kappa /3}}{q_{COBE}} & \in  & \left( 10^{-\frac \kappa 
2},\frac{\kappa}{2}\zeta \right) \left(\kappa-1\right)^{\frac{1-\kappa }2}
\left( 2\times 10^{18}GeV\right)^{2-\frac \kappa 2}
\end{array}
,
\]
where $\zeta \equiv 4\epsilon \left( \kappa -1\right) ^{\frac{\kappa -1}
\kappa }\simeq \frac{2(\kappa -1)10^{19}GeV}{\varphi _{cr}}\in (1,10)$;
recall these estimates assume only the condition T/S$>0.1$ (cf.eqs.(57),
(58), (68)).

\section{Conclusions}

Our conclusions are the following:

\begin{itemize}
\item[$\ast$]  We introduce a broad class of elementary inflaton models
called the $\Lambda $-inflation. The inflaton in the local-minimum-point has
a {\it positive residual potential energy}, $V_0>0$. The hybrid inflation
model (at the intermediate evolutionary stage) is a partial case of $\Lambda 
$-inflation; the chaotic inflation is a measure-zero-model in the family of 
$\Lambda $-inflation models.

\item[$\ast$]  The S-perturbation spectrum generated in $\Lambda $-inflation
has a non-power-law {\it wing-like shape with a broad minimum} where the
slope is locally HZ ($n_S=1$); it is blue, $n_S>1$, (red, $n_S<1$) on short
(large) scales. The T-perturbation spectrum remains always red with the
maximum deviation from HZ at the scale near the S-spectrum-minimum.

\item[$\ast$]  The cosmic gravitational waves generated in 
$\Lambda$-inflation contribute {\it maximally} to the SW 
$\Delta T/T$-anisortopy, (T/S)${}_{max}{}_{\sim }^{<}10$, in scales where 
the S-spectrum is slightly red or nearly HZ ($k_{\sim }^{<}k_{cr}$). The 
T/S remains small ($\ll 1$) in both blue ($k>k_{cr}$) and far red 
($k\ll k_{cr}$) S-spectrum asymptotics.

\item[$\ast$] {\it Three} independent arbitrary parameters determine the
fundamental $\Lambda $-inflation; they can be the T/S, $k_{cr}$ (the scale
where $n_S=1$), and $\sqrt{V_0}$ (the CDP amplitude at $k_{cr}$ scale; a
large value for T/S is expected if $V_0^{\frac 14}\sim 10^{16}GeV$). This 
brings high capability in fitting various observational tests to the dark 
matter cosmologies based on $\Lambda $-inflation.
\end{itemize}

{\it Acknowledgements} The work was partly supported by the INTAS grant 
97-1192 and Swiss National Science Foundation grant 7IP 050163.96/1.

\newpage

\section*{APPENDIX: $\Lambda m$-inflation with $\epsilon ^2<0.9$}

Here, we consider the inflation model with $\kappa =2$ and 
$p>\frac{2}{3}\;(\epsilon ^2<\frac 56)$.

Under the latter restriction, $\vartheta \simeq {\rm const}=\frac 32(1-p)$
during the whole dS stage ($y<1$, cf.eq.(42)) and decays as $\vartheta
\simeq \frac{3f}4\simeq -\frac{\epsilon ^2}{2y^2}$ for $y>1$. Making use of
eqs.(26) we find the following best fit for the whole $y$-evolution
(analytically exact in the limit $\epsilon \rightarrow 0$): 
$$
\vartheta \simeq \frac 32\left( 1-\sqrt{1-f}\right) =
\frac{1.5f}{1+\sqrt{1-f}},\eqno(A1)
$$
$$
\sqrt{\frac{\gamma}{2}}\simeq \frac{\epsilon y}{\left( 1+y^2\right) 
\left( 1+\sqrt{1-f}\right) }.\eqno(A2)
$$
where $f\equiv \frac{2\epsilon ^2}3\frac{1-y^2}{(1+y^2)^2}$. The
substitution of (A2) into eq.(27) brings about the explicit integration: 
$$
\epsilon ^2\ln \left( \frac {v\eta}{\sqrt{2}\eta_{cr}}\right) \simeq J(\xi ),
\eqno(A3)
$$
where $\xi \equiv v^2\left( 1+\sqrt{1-f}\right) =v^2+\sqrt{v^4+\left(
1-p^2\right) \left( v^2-2\right) }$, $v^2=1+y^2$, 
\[
J\left( \xi \right) \equiv \int_1\frac{\xi dy}y=\frac \xi 2-2+\frac 12\ln
\left[ \left( \frac{\xi -1-p}{3-p}\right) ^{1+p}\left( \frac{\xi -1+p}{3+p}
\right) ^{1-p}\left( \frac{2\xi +1-p^2}{9-p^2}\right) ^{\frac{1-p^2}2
}\right] .
\]

Obviously, the evolution goes from large $\xi =2y^2\left( 1+\frac{1+\epsilon
^2/6}{y^2}+O\left( \frac 1{y^4}\right) \right) >4$ to small $\xi =\left(
1+p\right) \left( 1+\frac{3-p}{2p}y^2+O\left( y^4\right) \right) <4$, and 
$\xi _{cr}=4$. Accordingly, we have the following $y$-asymptotics from
eq.(A3): 
$$
y^2=\Bigg \{
\begin{array}{lcl}
\epsilon ^2\ln \left( \frac{\omega _5\eta}{\eta_{cr}} \right) +\left( 1+p^2
\right)\ln \left( \frac{y_5}y\right) +1, & \; & y>1, \\ 
y_6^2\left( \frac{\omega _6\eta}{\eta_{cr}} \right) ^{3(1-p)}, & \; & y<1,  
\end{array}
\eqno(A4)
$$
where 
$\omega _5^{-1}=\sqrt{2}\exp \left( \frac 16\right)$, 
$y_5=\left(\frac{3-p}{2}\right)^{\frac{(1+p)(3-p)}{4(1+p^2)}}
\left(\frac{3+p}2\right)^{\frac{(1-p)(3+p)}{4(1+p^2)}}$, 
$\omega _6=\frac{\eta_{cr}}{\eta_3}=\frac 1{\sqrt{2}}(3+p)^{\frac{3+p}{6(1+p)
}}$, 
$y_6=(2p)^{\frac p{1+p}}(1+p)^{\frac{p-3}4}\exp\left[\frac{3-p}{2(1+p)}
\right] $.

In the allowed region $p>\frac 23$, the coefficients $\omega _6$ and $y_6$
remain close to unity. In the slow-roll limit $(p\rightarrow 1)$, $\omega
_6=2^{\frac 16}$ and $y_6=\sqrt{e}$.

\newpage

\section*{References}
1. E.M. Lifshitz, Zh. Eksp. Teor. Fiz. {\bf 16}, 587 (1946).\\
2. R.K. Sachs, A.M. Wolfe, ApJ {\bf 147}, 73 (1967).\\
3. L.P. Grishchuk, Zh. Eksp. Teor. Fiz. {\bf 67}, 825 (1974). \\
4. V.N. Lukash, Zh. Eksp. Teor. Fiz. {\bf 79}, 1601 (1980).\\
5. A.D. Linde, Phys. Lett. B {\bf 129}, 177 (1983).\\
6. F. Lucchin, S. Matarrese, Phys. Rev. D {\bf 32}, 1316 (1985).\\
7. R.L. Davis, H.M. Hodges, G.F. Smoot, et al., Phys. Rev. Lett. 
{\bf 69}, 1856 (1992).\\
8. A.D. Linde, Phys. Rev. D {\bf 49}, 748 (1994).\\
9. J. Garcia-Bellido, D. Wands, Phys. Rev. D {\bf 54}, 6040 (1996).\\
10. J. Garcia-Bellido, A. Linde, D. Wands, Phys. Rev. D {\bf 54}, 7181 
(1996).\\
11. E.J. Copeland, A.R. Liddle, D.H. Lyth, et al., Phys. Rev. D 
{\bf 49}, 6410 (1994).\\
12. V.N. Lukash, E.V. Mikheeva, Gravitation and Cosmology {\bf 2}, 
247 (1996).\\
13. B.J. Carr, J.H. Gilbert, Phys. Rev. D {\bf 50}, 4853 (1994).\\
14. J. Gilbert, Phys. Rev. D {\bf 52}, 5486 (1995). \\
15. A. Melchiorri, M.S. Sazhin, V.V. Shulga, N. Vittorio, to appear 
in ApJ, preprint astro-ph/9901220 (1999).\\
16. V.N. Lukash, in: {\it Cosmology: The physics of 
the Universe}, ed. by B.A. Robson et al., World Scientific, Singapore
(1996), p.213.\\
17. V.A. Rubakov, M.V. Sazhin, A.M. Veryaskin, Phys. Lett. B {\bf 
115}, 189 (1982).\\
18. A.A. Starobinskii, Zh. Eksp. Teor. Fiz. {\bf 30}, 719 (1979).\\
19. G.F. Smoot, C.L. Bennett, A.Kogut et al., ApJ {\bf 396}, L1 (1992).\\
20. C.L. Bennet, A.J. Banday, K.M.Gorski et al., ApJ {\bf 464}, L1 
(1996).\\
21. F. Lucchin, S. Matarrese, S. Mollerach, ApJ {\bf 401}, L49 (1992).\\
22. M.S. Turner, Phys. Rev. D {\bf 48}, 5539 (1993).\\
23. E.W. Kolb, S.L. Vadas, Phys.Rev.D {\bf 50}, 2479 (1994).\\
24. J.E. Lidsey, A.R.Liddle, E.W.Kolb et al., Rev. Mod. Phys. {\bf 69}, 373 
(1997).\\
25. A.A. Starobinsky, Pis'ma Astron.Zh. {\bf 11}, 323 (1985).\\
26. G. Lesgourgues, D. Polarski, A.A. Starobinsky, to appear in 
MNRAS, preprint astro-ph/9807019 (1999).

\end{document}